\documentclass[11pt,a4paper]{article}

\usepackage[utf8]{inputenc}
\usepackage[T1]{fontenc}
\usepackage{lmodern}
\usepackage{geometry}
\usepackage{microtype}
\usepackage{amsmath,amssymb,amsthm,mathtools}
\usepackage{bm} 
\usepackage{physics} 
\usepackage{graphicx}
\usepackage{caption}
\usepackage{subcaption}
\usepackage{tikz}
\usepackage{amsmath}
\usetikzlibrary{arrows.meta,patterns}
\usepackage{siunitx}
\usepackage[colorlinks=true, linkcolor=blue, urlcolor=blue, citecolor=blue]{hyperref}
\usepackage{cleveref}
\usepackage{enumitem}
\usepackage{booktabs}
\usepackage{natbib} 
\usepackage{graphicx}
\usepackage{subcaption}
\theoremstyle{plain}

\theoremstyle{definition}

\title{NON-EQUILIBRIUM MODELLING OF POLYATOMIC GASES:
GENERIC FRAMEWORK AND 11-MOMENT MODEL}

\author{
Masrakain Ahmad$^{1}$, Anil Kumar$^{1}$ Anirudh Singh Rana$^{1*}$\\
$^{1}$Department of Mathematics, Birla Institute of Technology and Science, Pilani,\\ Pilani Campus, Vidya Vihar, Pilani, Rajasthan 333031, India\\
*Corresponding Author: anirudh.rana@pilani.bits-pilani.ac.in 
}
\date{}
\begin{document}

\maketitle

\begin{abstract}
  In this study, we present a thermodynamically consistent framework for modelling polyatomic gases by integrating the \textit{general equation for non-equilibrium reversible-irreversible coupling} (GENERIC) with an 11-moment description. Our formulation unifies \textit{rational irreversible thermodynamics} (RIT) and \textit{extended irreversible thermodynamics} (EIT) into a single GENERIC-compatible structure that guarantees compliance with the second law of thermodynamics through the explicit construction of Poisson and friction matrices. By including translational and internal temperatures, as well as a temperature tensor, the model offers a well-structured and physically meaningful representation of non-equilibrium behavior in polyatomic gases. The entropy production is inherently non-negative, and all closure relations emerge naturally from thermodynamic principles. As a test of applicability, we compute the shock wave structure in nitrogen gas, validating the framework against Direct Simulation Monte Carlo (DSMC) simulations and experiments up to moderate Mach numbers.
\end{abstract}




\section{\label{sec:level1}Introduction}
The accurate modelling of polyatomic gases under non equilibrium conditions is essential for predicting the behavior of a wide range of physical systems, from aerospace re-entry flows to microscale heat transfer. Unlike monatomic gases, polyatomic gases possess internal degrees of freedom, such as rotational and vibrational modes that exchange energy with translational motion, resulting in rich relaxation dynamics and non-equilibrium effects \cite{myong1999, struchtrup2005,vincenti1965}. Capturing these interactions in a thermodynamically consistent and mathematically well-structured way remains a key challenge.

The classical Navier–Stokes–Fourier (NSF) equations are based on conservation laws with first-order constitutive relations \cite{struchtrup2005}, yielding a hyperbolic–parabolic structure that is mathematically tractable and easy to discretize. Moreover, they satisfy the second law of thermodynamics by ensuring non-negative entropy production, although they remain limited in capturing strong non-equilibrium effects and are accurate only to first order in the Knudsen number \cite{rahimi2016macroscopic}.
To overcome these limitations, extended models such as the coupled constitutive relations framework incorporate higher-order effects through coupled fluxes and gradients \cite{Anil2025CCR}. These models also satisfy the second law of thermodynamics and capture key rarefaction phenomena such as non-Fourier heat flux and thermal stresses. However, the coupling in the constitutive relations leads to a loss of the standard hyperbolic–parabolic structure, increasing the complexity of analysis and numerical implementation, while providing second-order accuracy in the Knudsen number.
The regularized Gaussian 11-moment (RG11) model, based on the Gaussian–gamma function, extends the moment framework by including additional non-equilibrium variables for translational and internal degree of freedom \cite{kumar2025regularized}. This results in a linearly stable system with improved description of rarefaction effects, achieving second-order accuracy comparable to advanced extended models. Despite these advantages, the RG11 model does not provide a closed-form entropy equation, which limits a rigorous assessment of thermodynamic consistency and entropy production. This motivates the development of the 11-moment model within the GENERIC framework, where the evolution equations are constructed to ensure a clear separation of reversible and irreversible processes and to guarantee consistency with the laws of thermodynamics. As a result, the proposed model provides a more complete and systematically derived thermodynamic description while retaining the strengths of the moment-based approach.

The GENERIC framework provides a unified thermodynamic foundation for describing the evolution of non-equilibrium systems. It combines both reversible (Hamiltonian) and irreversible (dissipative) dynamics in a structure-preserving manner, ensuring compatibility with the laws of thermodynamics. Specifically, the time evolution of the state variables $\mathcal{U} = \{\mathcal{U}_{\alpha}\}$ is governed by the GENERIC equation:
\begin{equation}
\frac{\partial \mathcal{U}_{\alpha}}{\partial t} = \mathcal{L}_{\alpha \beta}(\mathcal{U})\frac{\partial \mathcal{E}}{\partial \mathcal{U}_{\beta}} + \mathcal{M}_{\alpha \beta}(\mathcal{U})\frac{\partial \eta}{\partial \mathcal{U}_{\beta}},
\end{equation}
where $\mathcal{L}$ is the Poisson matrix encoding reversible dynamics, and $\mathcal{M}$ is the friction matrix responsible for dissipation. The functionals $\mathcal{E}(\mathcal{U})$ and $\eta(\mathcal{U})$ represent the total energy and entropy, respectively. 

In the context of polyatomic gases, the state vector $\mathcal{U}$ will include the mass density, the momentum density, and the distinct fields that represent the translational and internal energies. 

The Poisson matrix \(\mathcal{L}\) and the friction matrix \(\mathcal{M}\) must satisfy the degeneracy conditions \cite{ottinger2005beyond}
\begin{align}
    \mathcal{L}_{\alpha \beta}\frac{\partial \eta}{\partial \mathcal{U}_{\beta}} = 0, \quad \text{and} \quad \mathcal{M}_{\alpha \beta}\frac{\partial \mathcal{E}}{\partial \mathcal{U}_{\beta}} = 0.
\end{align}

The antisymmetry of $\mathcal{L}$ is defined through the Poisson bracket $\{A,B\}$,
\begin{align}
  \{A,B\}=\frac{\partial A }{\partial U_\alpha}\mathcal{L}_{\alpha \beta}\frac{\partial B}{\partial U_\beta},  
\end{align}
where $A(\mathcal{U})$ and $B(\mathcal{U})$ are arbitrary functionals of the variables $\mathcal{U}_{\alpha}$. The Poisson matrix has to obey the Jacobi identity
\begin{align}
    \{A,\{B,C\}\}+\{B,\{C,A\}\}+\{C,\{A,B\}\}=0.
\end{align}
The antisymmetry of $\mathcal{L}$, symmetry and positive semi-definiteness of $\mathcal{M}$, and the degeneracy conditions ensure the conservation of entropy under reversible dynamics and the conservation of energy under irreversible dynamics while entropy is non-decreasing \cite{grmela1997, ottinger1997, otto2005}.

The dissipative bracket associated with the friction matrix $\mathcal{M}$ is defined as
\begin{align}
    [A,B] = \frac{\partial A}{\partial \mathcal{U}_\alpha} \mathcal{M}_{\alpha \beta} \frac{\partial B}{\partial \mathcal{U}_\beta}.
\end{align}
If $\mathcal{M}$ is symmetric, then $[A,B] = [B,A]$, and non-negative entropy production requires
\begin{align}
    [A,A] \geq 0.
\end{align}
The GENERIC framework was originally formulated for complex fluids, with a comprehensive development provided by Öttinger \cite{ottinger2005beyond}. It was subsequently extended to solid mechanics through the work of Hütter and Svendsen \cite{hutter2011formulation,hutter2012thermodynamic}, as well as Mielke \cite{mielke2011formulation}.
 Romero \cite{romero2009thermodynamically} was the first to apply GENERIC to computational solid mechanics, introducing the concept of thermodynamically consistent integrators. More recently, Öttinger\cite{ottinger2018generic} proposed GENERIC integrators, which generalize symplectic integrators from Hamiltonian systems to include dissipative dynamics.

In the field of complex fluids and polymeric materials, the GENERIC framework has been instrumental in developing models that go beyond classical Newtonian rheology \cite{ottinger1999nonequilibrium}. Polymeric fluids exhibit memory effects, viscoelasticity, and anisotropy, all of which require a dynamic equation for internal structure, such as conformation tensors or microstructural variables. Beris and Edwards \cite{beris1994thermodynamics} formulation of polymer dynamics using a GENERIC-compatible structure allows consistent coupling between molecular configurations and macroscopic flow fields. In particular, the formulation of the Giesekus and FENE-P models using GENERIC ensures thermodynamic admissibility, enabling stable numerical simulations and better physical interpretation. These developments have not only refined our understanding of polymeric flows in extrusion and injection molding but have also proven essential in microfluidics and biomedical flows.

Furthermore, in solid mechanics and plasticity, GENERIC has been extended to model elasto-viscoplastic deformation under finite strains, incorporating internal variables such as defect densities or phase field order parameters. The strength of the framework lies in its capacity to simultaneously preserve geometric structures and incorporate dissipative mechanisms, a feature crucial for materials undergoing complex loading paths \cite{mielke2025general,betsch2019energy,hutter2012thermodynamic}.


Substantial applications of GENERIC arise in various domains where multiscale and non-equilibrium phenomena play a central role. One important direction is turbulence modelling. Traditional Reynolds-averaged and large eddy simulation approaches \cite{sagaut2006les} struggle with the closure problem and often rely on empirical assumptions. More recently, efforts have emerged to model turbulent flows using nonequilibrium thermodynamics, particularly exploiting the structure-preserving nature of GENERIC to capture energy cascades and entropy production. Eyink and Sreenivasan emphasize the importance of thermodynamic consistency in turbulence models, especially with regard to irreversibility and entropy \cite{eyink2006onsager}, while Öttinger and others have demonstrated that GENERIC provides systematic closure strategies based on entropy gradients, instead of purely phenomenological arguments \cite{ottinger1997}.

A further extension of GENERIC establishes links with rational extended thermodynamics (RET). The complete Boltzmann hierarchy aligns naturally with GENERIC however, truncating it while maintaining the Poisson structure is challenging  \cite{ottinger2020formulation}. Within this context, irreversible brackets—incorporating Onsager and Casimir symmetries—effectively represent dissipative dynamics and distinguish between entropy-producing and non-entropy-producing processes \cite{oettinger2014irreversible}. These insights are crucial for constructing thermodynamically consistent moment models.


The insights from previous thermodynamically consistent moment models directly inform our current work. In particular, the development of a 13-moment model for monoatomic gases \cite{struchtrup2022thermodynamically} with an explicit entropy satisfying an H-theorem and a Hamiltonian formulation for reversible transport provides a solid foundation. Additionally, the embedding of Grad’s 13-moment equations into the GENERIC framework, ensuring compatibility with the second law through Poisson and friction matrices, highlights the importance of maintaining thermodynamic structure. Building on these ideas, our work develops a GENERIC compatible 11-moment model for polyatomic gases that captures internal energy exchanges between translational, rotational, and vibrational modes while preserving both physical realism and mathematical consistency.


Building on these ideas, our work develops a GENERIC-compatible 11-moment model for polyatomic gases. The model captures internal energy exchanges between translational, rotational, and vibrational modes while preserving both physical realism and mathematical consistency. Specifically, we extend the classical moment method by explicitly including variables for translational and internal energies, their respective fluxes, and associated relaxation mechanisms. Closure is achieved via a RET-inspired formulation  \cite{muller1998, jou2010, ruggeri2015}, embedded within the GENERIC framework to ensure full compliance with thermodynamic laws. To the best of our knowledge, this is the first demonstration of a polyatomic gas model that combines GENERIC and RET-inspired moment closures in a thermodynamically consistent manner.


In fact, previous work embedding 13-moment equations into the GENERIC framework highlights the importance of preserving thermodynamic structure. The development of GENERIC-13 equations \cite{struchtrup2022thermodynamically}, for example, introduced a Poisson matrix for reversible dynamics and antisymmetric friction matrices to account for irreversible convective transport. This formulation guarantees compliance with the second law of thermodynamics while maintaining agreement with the classical Grad–13 equations up to second order in the Knudsen number. Similarly, the construction of a 13-moment model with an explicit entropy satisfying an H-theorem and a Hamiltonian formulation for reversible transport provides a strong foundation. Collectively, these thermodynamically admissible formulations motivate the extension of GENERIC-based approaches to higher-order moment systems.

The paper is organized as follows. Following this introduction, Section \ref{section kinetic theory} outlines the kinetic theory of polyatomic gases. The conservation laws and extended balance equations are presented in Section \ref{section conservation laws}, while Section \ref{sec of summery of model} provides a summary of the proposed 11-moment model. Section \ref{section GENERIC} is dedicated to the GENERIC formulation, where the thermodynamic structure of the model is established. In Section \ref{section shock structure}, the proposed model is employed to investigate the shock wave structure in nitrogen gas, and the obtained results are compared with those of the two-temperature (2T) model \cite{kumar2024capturing}, as well as with available experimental measurements \cite{alsmeyer1976density} and DSMC simulations \cite{cai2014nrxx}. Finally, Section \ref{Section Summary and future direction} offers a detailed discussion of the results along with concluding remarks and future research directions.
\section{\label{section kinetic theory}Kinetic theory of polyatomic gases}

The dynamics of polyatomic gases are governed by the Boltzmann equation for the single-particle distribution function $f(t,x_i,c_i,I)$, given by
\begin{equation}
\frac{\partial f}{\partial t}+c_{k}\frac{\partial f}{\partial x_{k}}+F_{k}\frac{\partial f}{\partial c_{k}}=\mathcal{S}[f,f],
\label{Boltzman equation}
\end{equation}
where $f$ denotes the distribution function of gas molecules with three translational and internal degrees of freedom $\delta$. The variables are: time $t$, spatial position $x_i \in \mathbb{R}^3$, molecular velocity $c_i \in \mathbb{R}^3$, $F_k$ is an external force (e.g. gravity, often neglected in many gas dynamics problems), and internal energy $I \in \mathbb{R}^+$. The internal energy is treated classically, neglecting quantization, allowing $\delta$ to be non-integer and potentially temperature-dependent. Einstein summation convention is adopted throughout.

The right-hand side of (\ref{Boltzman equation}), denoted $\mathcal{S}[f,f]$, is the collision operator, representing changes in $f$ due to binary molecular collisions. This operator typically involves complex integrals depending on the intermolecular potential. Nevertheless, it conserves five fundamental quantities, known as the collision invariants,  which are given by:

\begin{equation}
\psi =m\left\{ 1\text{, }c_{i}\text{, }\frac{C^{2}}{2}+I\right\}\text{,}
\label{collision invariants}
\end{equation}
i.e., $\int \mathcal{S}\psi dcdI=0$.
 The macroscopic quantities such as mass density $\rho$, momentum density $\rho v_i$, and total energy $u$, are obtained as moments of the distribution function $f$, and are defined as follows:
\begin{subequations}
\begin{eqnarray}
&\rho = m\int fdcdI, \quad
\rho v_{i} = m\int c_{i}fdcdI, \label{eq:rho_vi} \\
&\rho u = m\int \left( \frac{C^{2}}{2}+I\right) fdcdI\text{,} \label{eq:rho_u}
\end{eqnarray}
\end{subequations}
where $m$ is the molecular mass, and $C_i$ $=c_i-v_i$ is the
peculiar velocity and $v_i$ is the centre of mass velocity. 
Furthermore, the total energy $u$, can be divided into
two parts: (i) the translational part $u^{tr}$ and (ii) part due to the
internal degrees of freedom $u^{in}$, as%
\begin{subequations}
  \begin{eqnarray}
\rho u^{tr} &:=&\frac{3}{2}\rho \theta _{tr}=m\int f\frac{C^{2}}{2}dcdI\text{, }  \label{moments translational energy} \\
\rho u^{in} &:=&\frac{\delta }{2}\rho \theta _{in}=m\int fIdcdI\text{.}  \label{moments internal energy}
\end{eqnarray}%
\label{combined moments}
\end{subequations}
Here, we introduce two distinct temperature components: the translational temperature $\theta_{\text{tr}}$, associated with the kinetic energy of molecular motion, and the internal temperature $\theta_{\text{in}}$, associated with internal degrees of freedom (e.g., rotation or vibration). Both temperatures are expressed in energy units. The thermodynamic temperature $\theta = \mathrm{R}T$, where $\mathrm{R}=k_b/m$ is the specific gas constant, $k_b$ is the Boltzmann constant, and $T$ is the absolute temperature in Kelvin, is related to these two components by the following relation
\begin{equation}
\frac{3+\delta }{2}\theta =\frac{3}{2}\theta _{tr}+\frac{\delta }{2}\theta
_{in}\text{.}  \label{temperature}
\end{equation}%
This equation ensures energy equipartition among translational and internal modes in equilibrium.

The entropy density is defined by the relation \cite{Rana2014}
\begin{equation}
 \eta=\rho s=-k_b\int f\ln \frac{f}{I^{\delta /2-1}}dcdI\text{,}  \label{Entropy functional}
\end{equation}
where $s$ denotes the specific entropy of the gas and the entropy law written as
\begin{align}
    \rho \frac{Ds }{Dt}+\frac{\partial \left( \frac{q_{k}^{in}}{\theta _{in}}+%
\frac{s_{k}}{\theta _{tr}}\right) }{\partial x_{k}} =\sigma_{p}\label{Enrtopy Law},
\end{align}
where $\sigma_{p}$ is entropy production rate.
\\
\underline{\textit{Proposition 1}}: \textit{The distribution which maximizes the entropy (\ref{Entropy functional})
under six constraints (\ref{eq:rho_vi}), and (\ref{combined moments}) takes the
following form 
\begin{equation}
f_{\mathrm{6}}=\underset{\text{Maxwellian}}{\frac{\rho }{m}\underbrace{\frac{%
1}{\sqrt{2\pi \theta _{tr}}^{3}}e^{-\frac{C^{2}}{2\theta _{tr}}}}}\underset{%
\text{Gamma}}{\underbrace{\frac{1}{\Gamma \left( \frac{\delta }{2}\right) }%
\frac{1}{I}\left( \frac{I}{\theta _{in}}\right) ^{\delta /2}e^{-\frac{I}{%
\theta _{in}}}}}.
\end{equation}
}
The proof of this proposition is given in \cite{kumar2023h}, with additional details in the Appendix.
Furthermore, one defines the temperature tensor $\rho\Theta_{ij}$ as
\begin{equation}
\rho \Theta_{ij} := m\int fC_{i}C_{j}dcdI,
\label{moments translational temperature tensor}
\end{equation}
so that (\ref{moments translational energy}) gives $\Theta_{kk}=3\theta_{tr}$ as the trace of $\Theta$; $\Theta_{ij}$ is the symmetric and positive-definite covariance matrix.\\
\underline{\textit{Proposition 2}}:
\textit{The distribution which maximizes the entropy (\ref{Entropy functional})
under the eleven constraints (\ref{eq:rho_vi}), (\ref{combined moments}), and (\ref{moments translational temperature tensor})
takes the following form%
\begin{equation}
f_{\mathrm{11}}=\underset{\text{Gaussian}}{\frac{\rho }{m}\underbrace{\frac{1%
}{\sqrt{\det \left( 2\pi \Theta \right) }}e^{-\frac{1}{2}\Theta
_{ij}^{-1}C_{i}C_{j}}}}\underset{\text{Gamma}}{\underbrace{\frac{1}{\Gamma
\left( \frac{\delta }{2}\right) }\frac{1}{I}\left( \frac{I}{\theta _{in}}%
\right) ^{\delta /2}e^{-\frac{I}{\theta _{in}}}}}.
\label{11 moment distribution}
\end{equation}
}
A detailed proof of this result can be found in \cite{kumar2025regularized} (see also the Appendix therein).
The entropy expression for the 11-moment system is derived by substituting $f_{11}$ from equation (\ref{11 moment distribution}) into the entropy functional (\ref{Entropy functional}). This leads to the formulation of an extended Gibbs relation as follows:
\begin{equation}
ds = \frac{\delta}{2} \frac{1}{\theta_{in}} d\theta_{in} + \frac{1}{2}\Theta_{ij}^{-1}d\Theta_{ij} - \frac{1}{\rho}d\rho.
\label{extended Gibbs relation}
\end{equation}
\section{The conservation laws and the extended balance equations} \label{section conservation laws}
The conservation laws are derived from the Boltzmann equation (\ref{Boltzman equation}) by multiplying it with the collision invariants (\ref{collision invariants}) and integrating over velocity space $c_i$ and internal energy $I$, yielding
\begin{subequations}
\label{Conservations laws}
\begin{eqnarray}
\frac{D\rho }{Dt}+\rho \frac{\partial v_{k}}{\partial x_{k}} &=&0\text{,}
\label{mass  conservation} \\
\rho \frac{Dv_{i}}{Dt}+\frac{\partial \rho \Theta _{ik}}{\partial x_{k}} &=&0%
\text{,}  \label{momentum conservation} \\
\rho \frac{D\left( \frac{3+\delta }{2}\theta \right) }{Dt}+\rho \Theta _{kr}%
\frac{\partial v_{r}}{\partial x_{k}}+\frac{\partial q_{k}}{\partial x_{k}}
&=&0\text{.}  \label{energy conservation}
\end{eqnarray}%
\label{Conservation equation}
\end{subequations}
Here, D/Dt denotes the convective time derivative, $q_k$ is the total heat flux.
The balance equations for the temperature tensor, internal and translational temperature are obtained as 
\begin{eqnarray}
\rho \frac{D\Theta _{ij}}{Dt}+\rho \Theta _{ik}\frac{\partial v_{j}}{\partial x_{k}}+\rho \Theta _{jk}\frac{\partial v_{i}}{\partial x_{k}}+\frac{\partial \varrho _{ijk}^{0,0}}{\partial x_{k}} &=&\mathcal{P}_{ij}^{0,0}, \label{temperature tensor equation 1} \\
\rho \frac{D\left( \frac{\delta }{2}\theta _{in}\right) }{Dt}+\frac{\partial q_{k}^{in}}{\partial x_{k}} &=&\mathcal{P}^{0,1} \label{internal temperature},\\
\frac{3}{2}\rho\frac{D}{Dt}\theta_{tr}+\rho \Theta_{kr}\frac{\partial v_r}{\partial x_k}+\frac{\partial q_k^{tr}}{\partial x_k}&=&-\mathcal{P}^{0,1}.\label{Translation temp eqn}
\end{eqnarray}
 Here, we have introduced the internal heat flux $q_{k}^{in}$, translational heat flux $q_{k}^{tr}$, the total heat flux $q_{k}$, and a third order moment $\varrho _{ijk}^{0,0}$, which are defined as 
\begin{eqnarray}
q_{k}^{in} &:=&m\int fIC_{k}dcdI\text{, } \quad q_{k}^{tr}:=m\int f\frac{C^{2}}{2}C_{k}dcdI\text{, } \\
q_{k} &:=&q_{k}^{tr}+q_{k}^{in}\text{, }\quad \varrho _{ijk}^{0,0}:=m\int fC_{i}C_{j}C_{k}dcdI\text{.}
\end{eqnarray}
Hence, $\varrho _{ikk}^{0,0}=2q_{i}^{tr}$. The production terms $\mathcal{P}_{ij}^{0,0}$ and $\mathcal{P}^{0,1}$ in equations (\ref{temperature tensor equation 1}) and (\ref{internal temperature}) stem from the Boltzmann collision operator, so that 
\begin{equation}
\mathcal{P}_{ij}^{0,0}:=m\int \mathcal{S}C_{i}C_{j}dcdI\text{, and }\mathcal{P}^{0,1}:=m\int \mathcal{S}IdcdI\text{,}
\end{equation}
since $\frac{C^{2}}{2}+I$ is a collision invariant, we have $\mathcal{P}_{kk}^{0,0}=-2\mathcal{P}^{0,1}$.\\
\underline{\textit{Proposition 3}}:
\textit{Any symmetric third-order tensor $A_{ijk}$ (in three dimensions) can be expressed as:%
\begin{equation}
A_{ijk}=\left( A_{i}^{(1)}\Theta _{jk}+A_{j}^{(1)}\Theta _{ik}+A_{k}^{(1)}\Theta _{ij}\right) +A_{\left\{ ijk\right\} },
\end{equation}%
where $\Theta $ represents any given symmetric positive definite matrix, and the anisotropic part (with respect to $\Theta $), $A_{\left\{ ijk\right\} }$, is a symmetric third-order tensor satisfying: 
\begin{equation*}
A_{\left\{ ijk\right\} }\Theta _{ik}^{-1}=A_{\left\{ ijk\right\} }\Theta _{jk}^{-1}=A_{\left\{ ijk\right\} }\Theta _{ij}^{-1}=0.
\end{equation*}%
Furthermore, $A_{i}^{(1)}=\frac{1}{5}A_{ijk}\Theta _{jk}^{-1}$. It is evident that if $\Theta $ is an identity matrix, then $A_{\left\{ ijk\right\} }$ becomes the trace-free component of $A_{ijk}$.}\\
For convenience, let us write the third-order moment 
\begin{equation}
\varrho _{ijk}^{0,0}:=\frac{2}{5\theta _{tr}}\left( s_{i}\Theta _{jk}+s_{j}\Theta _{ik}+s_{k}\Theta _{ij}\right) +\varrho _{\left\{ ijk\right\} }^{0,0}\text{.}  \label{third order moment}
\end{equation}%
Here, we have introduced the vector $s_i$, having the same dimensions as the heat flux, we shall show later that it is a more natural choice as a variable and constitutes the entropy flux. Nevertheless, the following equation establishes the relationship between the translational heat flux $q_{i}^{tr}$ 
\begin{equation}
q_{i}^{tr}=\frac{\varrho _{ikk}^{0,0}}{2}=\left( \delta _{ik}+\frac{2}{5\theta _{tr}}\Theta _{\left\langle ik\right\rangle }\right) s_{k}+\frac{\varrho _{\left\{ ikk\right\} }^{0,0}}{2}  \label{heat flux and s},
\end{equation}%
where $\Theta_{\left\langle ij\right\rangle} = \Theta_{ij}-\theta_{tr}\delta_{ij}$ is the traceless part of $\Theta_{ij}$, and it is related to the stress tensor via the relation $\sigma_{ij}=\rho \Theta_{\left\langle ij\right\rangle}=\rho \theta_{ij}$. It is easy to verify that the matrix $\left(\delta_{ij} + \frac{2}{5\theta_{tr}}\Theta_{\left\langle ij\right\rangle}\right)$ is invertible as well as positive definite.

Introducing  third-order moment $\varrho _{ijk}^{0,0}$ from the equation (\ref{third order moment}) in equation (\ref%
{temperature tensor equation 1}), we get%
\begin{align}
\rho \frac{D\Theta _{ij}}{Dt} & + \rho \Theta _{ik}\frac{\partial v_{j}}{\partial x_{k}} + \rho \Theta _{jk}\frac{\partial v_{i}}{\partial x_{k}}  + \frac{2}{5}\frac{\partial \left( \frac{s_{i}\Theta _{jk}+s_{j}\Theta_{ik}+s_{k}\Theta _{ij}}{\theta _{tr}}\right) }{\partial x_{k}} + \frac{\partial \varrho _{\left\{ ijk\right\} }^{0,0}}{\partial x_{k}} = \mathcal{P}_{ij}^{0,0}. \label{temperature tensor equation 2}
\end{align}

Indeed, in order to close the governing equations (\ref{Conservations laws}, \ref{internal temperature}, \ref{temperature tensor equation 2}), constitutive relations for internal heat flux \( q_{k}^{in} \), \( s_{i} \), and \( \varrho_{\{ijk\}}^{0,0} \) need to be postulated, which we shall propose through the validity of the second law inequality in the next section.

Using the extended Gibbs equation (\ref{extended Gibbs relation}), along with conservation laws (\ref{Conservations laws}), and the balance equations (\ref{internal temperature}) and (\ref%
{temperature tensor equation 2}), yields the time
rate of change of the entropy of a material element as
\begin{align}
\rho \frac{Ds }{Dt}+\frac{\partial \left( \frac{q_{k}^{in}}{\theta _{in}}+%
\frac{s_{k}}{\theta _{tr}}\right) }{\partial x_{k}} =-\frac{1}{\theta
_{in}^{2}}q_{k}^{in}\frac{\partial \theta _{in}}{\partial x_{k}}+\frac{1}{2}\Theta _{ij}^{-1}\mathcal{P}_{ij}^{0,0}+\frac{1%
}{\theta _{in}}\mathcal{P}^{0,1}\nonumber \\
+\frac{1}{%
5\theta _{tr}}s_{i}\Theta _{i\alpha }^{-1}\left[ \Theta _{l\alpha }\Theta
_{jk}+\Theta _{lj}\Theta _{\alpha k}+\Theta _{lk}\Theta _{\alpha j}\right] 
\frac{\partial \Theta _{jk}^{-1}}{\partial x_{l}}\nonumber \\
-\frac{1}{2}\varrho _{\left\{ ijk\right\} }^{0,0}\Theta _{i\alpha
}^{-1}\Theta _{j\beta }^{-1}\Theta _{k\gamma }^{-1}A_{\left\{ \alpha \beta
\gamma \right\} },\label{Entropy equation}
\end{align}
where
\begin{equation*}
A_{ijk}=\left( \Theta _{li}\frac{\partial \Theta _{jk}}{\partial x_{l}}%
+\Theta _{lj}\frac{\partial \Theta _{ik}}{\partial x_{l}}+\Theta _{lk}\frac{%
\partial \Theta _{ij}}{\partial x_{l}}\right) \text{.}
\end{equation*}%
After comparing equation (\ref{Entropy equation}) with the entropy law (\ref{Enrtopy Law}), the right-hand side of equation (\ref{Entropy equation}) is identified as the entropy production rate
\begin{align}
    \sigma_{p}=-\frac{1}{\theta
_{in}^{2}}q_{k}^{in}\frac{\partial \theta _{in}}{\partial x_{k}}
+\frac{1}{%
5\theta _{tr}}s_{i}\Theta _{i\alpha }^{-1}\left[ \Theta _{l\alpha }\Theta
_{jk}+\Theta _{lj}\Theta _{\alpha k}+\Theta _{lk}\Theta _{\alpha j}\right] 
\frac{\partial \Theta _{jk}^{-1}}{\partial x_{l}} \nonumber\\
-\frac{1}{2}\varrho _{\left\{ ijk\right\} }^{0,0}\Theta _{i\alpha
}^{-1}\Theta _{j\beta }^{-1}\Theta _{k\gamma }^{-1}A_{\left\{ \alpha \beta
\gamma \right\} }+\frac{1}{2}\Theta _{ij}^{-1}\mathcal{P}_{ij}^{0,0}+\frac{1%
}{\theta _{in}}\mathcal{P}^{0,1}.
\end{align}
We take the higher-order moment $ \varrho _{\left\{ ijk\right\} }^{0,0}$ is zero for simplicity in the rest of the paper \cite{rahimi2016macroscopic}. However, the same procedure can also be applied when this moment is non-zero.

According to the second law of thermodynamics, the entropy production rate must be non-negative, ensuring that entropy increases over time. To satisfy this requirement, it is sufficient to impose constitutive equations of the form
\begin{align}
      s_i=\frac{\gamma_1 }{5\theta_{tr}}[\Theta_{li}\Theta_{jk}+\Theta_{lj}\Theta_{ik}+\Theta_{lk}\Theta_{ij}]\frac{\partial \Theta_{jk}^{-1}}{\partial x_l}+\gamma_1\Bar{\omega} \Theta_{ik}\frac{\partial \theta_{in}^{-1}}{\partial x_k},
      \label{eqn of si}
\end{align}
\begin{align}
     q_i^{in}=\frac{\Bar{\omega}\gamma_1}{5\theta_{tr}}[\Theta_{li}\Theta_{jk}+\Theta_{lj}\Theta_{ik}+\Theta_{lk}\Theta_{ij}]\frac{\partial \Theta_{jk}^{-1}}{\partial x_l}
    +[\gamma_1\Bar{\omega}^2+\gamma_2]\Theta_{ik}\frac{\partial \theta_{in}^{-1}}{\partial x_k}\label{Qin}, 
\end{align}
\begin{align}
      q_i^{tr} = \left( \delta_{ik} + \frac{2}{5 \theta_{tr}} \Theta_{\langle ik \rangle} \right) s_k =\left(\frac{2}{5\theta_{tr}}\Theta_{ik}+\frac{3}{5}\delta_{ik}\right)s_k\label{Qtr},
\end{align}
The coefficient multiplying the internal-temperature force in \eqref{Qin} is chosen to ensure that the heat-flux contribution to the entropy production remains non-negative. For simplicity, let $X_1$ denote the thermodynamic force related to the translational temperature tensor, and let $X_2$ denote the force associated with the gradient of the internal temperature. Then, the corresponding coupled flux--force relation can be written as
\begin{equation}
\begin{pmatrix}
J_1\\
J_2
\end{pmatrix}
=
\begin{pmatrix}
\gamma_1 & \gamma_1\bar{\omega}\\
\gamma_1\bar{\omega} & \gamma_1\bar{\omega}^2+\gamma_2
\end{pmatrix}
\begin{pmatrix}
X_1\\
X_2
\end{pmatrix}.
\label{eq:onsager_heat_block}
\end{equation}
The corresponding entropy-production contribution is
\begin{equation}
J_1X_1+J_2X_2
=
\gamma_1\left(X_1+\bar{\omega}X_2\right)^2
+
\gamma_2X_2^2
\geq 0,
\label{eq:positive_heat_block}
\end{equation}
provided that $\gamma_1\geq0$ and $\gamma_2\geq0$. Thus,
$\bar{\omega}$ allows the coupling between translational and internal degree of freedom, while $\gamma_1,\gamma_2$ are another positive phenomenological coefficients.

 The choice of the production term $\mathcal{P}^{0,1}$ and $\mathcal{P}_{\langle
     ij \rangle}^{0,0}$ is based on the comparison with the model presented in \cite{kumar2023h,rahimi2016macroscopic}
\begin{align}
\mathcal{P}^{0,1}=\frac{\delta\rho^2\vartheta \theta}{(3+\delta)\mu_b
    },\quad
     \mathcal{P}_{\langle
     ij \rangle}^{0,0}=-\frac{\rho^2\theta}{\mu }\Theta_{\langle ij \rangle}.
\end{align}
Here, $\vartheta$ denotes the non-equilibrium contribution to the temperature, commonly referred to as the dynamic temperature, and is defined as $\vartheta = \theta_{tr} - \theta$ \cite{kumar2024capturing}. Furthermore, $\mu_b$ represents the bulk viscosity, while  $\mu$ denotes the shear viscosity of the gas.

By taking the trace part of equations \eqref{Qin} and \eqref{Qtr} by using \eqref{eqn of si}, (see Appendix \ref{Appendix for derivation1}), we get \eqref{trace part of eqn qtr qin}, and comparing it with the equation (28) from the \cite{kumar2023h}, we derive the following relationships
\begin{align}
    \gamma_1 = \frac{\zeta_{11}}{\theta_{tr}}, \quad \gamma_2 = \frac{1}{\theta_{tr}} \left( \zeta_{22} - \frac{\zeta_{12}^2}{\zeta_{11}} \right), \quad \Bar{\omega} = \frac{\zeta_{12}}{\zeta_{11}}.
\end{align}
Here, the values of the arbitrary non-negative coefficients $\zeta_{11}$, $\zeta_{12}$, and $\zeta_{22}$ are determined by comparison with various models available in the literature \cite{aoki2020two,djordjic2023boltzmann,marques1993spectral,rahimi2016macroscopic}. It is important to note that these phenomenological coefficients can vary depending on the specific collision model adopted. In this work, we follow the model proposed by Marques and Kremer \cite{marques1993spectral}, who developed a hydrodynamic framework for polyatomic gases with spherical molecules, incorporating rotational energy. Based on their formulation, the coefficients are given by \cite{kumar2023h,marques1993spectral}
\begin{align}
\zeta_{11} &= \dfrac{15\mu(6 + 25\varsigma + 38\varsigma^{2} + 26\varsigma^3)}{24 + 150\varsigma + 202\varsigma^{2} + 204\varsigma^3}\theta_{tr}^2, \nonumber\\
\zeta_{12} &= \dfrac{15\mu(6 + 13\varsigma)\varsigma}{24 + 150\varsigma + 202\varsigma^{2} + 204\varsigma^3}\theta_{in}^2, \nonumber\\
\zeta_{22} &= \dfrac{3\theta_{in}^2(6+13\varsigma)(25\theta_{in}^2\varsigma +2\theta_{tr}^2(6+13\varsigma))\mu}{10\theta_{tr}^2(12 + 75\varsigma + 101\varsigma^{2} + 102\varsigma^3)}.
\end{align}
Here, $\varsigma$ is the dimensionless moment-of-inertia parameter defined as,
\begin{equation}
\varsigma=\frac{4I}{ma^2},
\end{equation}
where $I$ is the molecular moment of inertia, $m$ is the molecular mass, and $a$
is the molecular diameter. With this convention, a uniform solid sphere has
$I=ma^2/10$, whereas a hollow spherical shell has $I=ma^2/6$. Therefore
$\varsigma\in[0,2/3]$. The value $\varsigma=0$ corresponds to negligible rotational
inertia, while $\varsigma=2/3$ corresponds to the limiting value associated with a
surface-mass distribution \cite{gaio1991kinetic,rodbard1990kinetic}.

\section{Theoretical Properties of the 11-Moment Model}\label{sec of summery of model}
The model consists of a closed system of extended balance equations derived for polyatomic gases within a thermodynamically admissible framework. It includes conservation equations for mass (\ref{mass eqn}), momentum (\ref{momentum eqn}), and total energy (\ref{energy eqn}), as well as separate evolution equations for internal energy (\ref{internal energy eqn}) and the second-order temperature tensor (\ref{translational temp eqn}), which captures non-equilibrium stress contributions. These equations incorporate both translational and internal degrees of freedom, along with corresponding heat fluxes and higher-order moment interactions. The system is closed using constitutive relations, which are related to the translational heat flux  (\ref{trans. heat flux}) and internal heat flux $q_{k}^{in} $  (\ref{internal heat flux eqn}).
\begin{subequations}
\begin{eqnarray}
\frac{D\rho }{Dt}+\rho \frac{\partial v_{k}}{\partial x_{k}} &=&0\text{,}\label{mass eqn} \\
\rho \frac{Dv_{i}}{Dt}+\frac{\partial \rho \Theta _{ik}}{\partial x_{k}} &=&0%
\text{,}  \label{momentum eqn} \\
\rho \frac{D\left( \frac{3+\delta }{2}\theta \right) }{Dt}+\rho \Theta _{kr}%
\frac{\partial v_{r}}{\partial x_{k}}+\frac{\partial q_{k}}{\partial x_{k}}
&=&0\text{,} \label{energy eqn}\\
\rho \frac{D\left( \frac{\delta }{2}\theta _{in}\right) }{Dt}+\frac{\partial q_{k}^{in}}{\partial x_{k}} &=&\mathcal{P}^{0,1} \label{internal energy eqn},\\
\rho \frac{D\Theta _{ij}}{Dt}  + \rho \Theta _{ik}\frac{\partial v_{j}}{\partial x_{k}} + \rho \Theta _{jk}\frac{\partial v_{i}}{\partial x_{k}}  + \frac{2}{5}\frac{\partial \left( \frac{s_{i}\Theta _{jk}+s_{j}\Theta_{ik}+s_{k}\Theta _{ij}}{\theta _{tr}}\right) }{\partial x_{k}} &=& \mathcal{P}_{ij}^{0,0}. \label{translational temp eqn}
\end{eqnarray}%
\label{Conservation and balance equations}
\end{subequations}
To close the above system, we use the constitutive equations of heat fluxes,
\begin{subequations}
    \begin{align}
      s_i=\frac{\gamma_1 }{5\theta_{tr}}[\Theta_{li}\Theta_{jk}+\Theta_{lj}\Theta_{ik}+\Theta_{lk}\Theta_{ij}]\frac{\partial \Theta_{jk}^{-1}}{\partial x_l}+\gamma_1\Bar{\omega} \Theta_{ik}\frac{\partial \theta_{in}^{-1}}{\partial x_k}\label{trans. heat flux}
\end{align}
\begin{align}
     q_i^{in}=\frac{\Bar{\omega}\gamma_1}{5\theta_{tr}}[\Theta_{li}\Theta_{jk}+\Theta_{lj}\Theta_{ik}+\Theta_{lk}\Theta_{ij}]\frac{\partial \Theta_{jk}^{-1}}{\partial x_l}
    +[\gamma_1\Bar{\omega}^2+\gamma_2]\Theta_{ik}\frac{\partial \theta_{in}^{-1}}{\partial x_k}\label{internal heat flux eqn}
\end{align}
\begin{align}
      q_i^{tr} = \left( \delta_{ik} + \frac{2}{5 \theta_{tr}} \Theta_{\langle ik \rangle} \right) s_k =\left(\frac{2}{5\theta_{tr}}\Theta_{ik}+\frac{3}{5}\delta_{ik}\right)s_k
\end{align}
\label{Constitutive eqn}
\end{subequations}
To further understand the behavior of the proposed 11-moment model, we perform an order-of-magnitude analysis in the following section \cite{rahimi2016macroscopic, struchtrup2005}. Under appropriate scaling assumptions, the system reduces to the Euler equations at leading order and to the NSF equations at the next order.
\subsection{Reduction to the Euler Limit}
For the hydrodynamic limit of the 11-moment system, we first define the variables according to their asymptotic orders \cite{rahimi2016macroscopic,struchtrup2005}
\begin{subequations}
\begin{align}
\mathcal{O}(\epsilon^0) &: \rho, v_i, \theta,\\
\mathcal{O}(\epsilon^\alpha) &: \vartheta,\\
\mathcal{O}(\epsilon^1) &: \sigma_{ij}, q_i.
\end{align}
\end{subequations}
Accordingly, the scaled equations \eqref{mass eqn}, \eqref{momentum eqn}, and \eqref{energy eqn} are rewritten as 
\begin{subequations}
    \begin{align}
\frac{D\rho}{Dt} + \rho \frac{\partial v_k}{\partial x_k} &= 0, \\[6pt]
\frac{D v_i}{Dt} + \theta \frac{\partial \ln \rho}{\partial x_i} + \frac{\partial \theta}{\partial x_i}
+ \epsilon^{\alpha} \left[ \frac{\partial  \vartheta}{\partial x_i} +  \vartheta \frac{\partial \ln \rho}{\partial x_i} \right]
+ \epsilon^1 \left[ \frac{1}{\rho} \frac{\partial \sigma_{ik}}{\partial x_k} \right] &= 0, \\[6pt]
\frac{3+\delta}{2} \rho \frac{D\theta}{Dt}
+ \rho \theta \frac{\partial v_k}{\partial x_k}
+ \epsilon^{\alpha} \left[ \rho  \vartheta \frac{\partial v_k}{\partial x_k} \right]
+ \epsilon^1 \left[ \frac{\partial q_k}{\partial x_k} + \sigma_{ik} \frac{\partial v_k}{\partial x_i} \right] &= 0.
\end{align}\label{scaled equations}
\end{subequations}
We begin the reduction process by retaining only the zeroth-order terms in \eqref{scaled equations} then the system reduces to the well-known classical Euler equations of gas dynamics 
\begin{subequations}
    \begin{align}
\frac{D \rho}{D t} + \rho \frac{\partial v_k}{\partial x_k} &= 0, \\
\frac{D v_i}{D t}  + \frac{\partial \theta}{\partial x_i} +\frac{\theta}{\rho} \frac{\partial \rho}{\partial x_i}&= 0, \\
\frac{3 + \delta}{2}\, \rho \frac{D \theta}{D t} + \rho \theta \frac{\partial v_k}{\partial x_k} &= 0.
\end{align}
\end{subequations}
The resulting system of equations constitutes a closed set governing the variables $\rho$, $v_i$, and $\theta$.

For the first-order approximation, terms up to order $\mathcal{O}(\epsilon^1)$ are retained in the conservation laws \eqref{scaled equations} and constitutive equations \eqref{Constitutive eqn}, implying that all contributions in the governing equations become relevant at this order. For $\alpha < 0.5$, the next-order contributions arise at order $\epsilon^{2\alpha}$. Since $\sigma_{ij}$ and $q_i$ are of order $\epsilon^1$, they remain negligible at this stage. However, the higher-order contributions of $\vartheta$ must be retained, which provides the closure through the full balance equation for $\vartheta$. Consequently, the system reduces to the two-temperature NSF equations \cite{kumar2023h,rahimi2016macroscopic}, governed by the independent field variables  $\{\rho, v_i, \theta, \vartheta\}.$
\begin{align}
\frac{D\rho}{Dt} + \rho \frac{\partial v_i}{\partial x_i} &= 0, \\[6pt]
\frac{D v_i}{Dt} + \theta \frac{\partial \ln \rho}{\partial x_i} + \frac{\partial \theta}{\partial x_i}
+ \frac{\partial \vartheta}{\partial x_i} + \vartheta \frac{\partial \ln \rho}{\partial x_i}
+ \frac{1}{\rho} \frac{\partial \sigma_{ij}}{\partial x_j} &= 0, \\[6pt]
\frac{3+\delta }{2} \rho \frac{D\theta}{Dt}
+ \rho \theta \frac{\partial v_i}{\partial x_i}
+ \rho \vartheta \frac{\partial v_i}{\partial x_i}
+ \frac{\partial q_i}{\partial x_i} + \sigma_{ij} \frac{\partial v_j}{\partial x_i} &= 0,\\[6pt]
 \rho \frac{D\vartheta}{Dt} + \frac{2\delta}{3(3+\delta)}\rho (\theta+\vartheta) \frac{\partial v_i}{\partial x_i} &= -\frac{2}{3}\mathcal{P}^{0,1},
\end{align}
and constitutive equation,
\begin{align}
    \sigma_{ij} = -2\mu \frac{\partial v_{\langle i}}{\partial x_{j\rangle}}, \qquad q_k = -\kappa \frac{\partial \theta}{\partial x_k},
    \label{Reduced NSF constitutive eqn}
\end{align}
where the heat conductivity is defined as $\kappa=\frac{\zeta_{11}+2\zeta_{12}+\zeta_{22}}{\theta^2}$.
\subsection{Mathematical structure}
The governing equations can be written in the conservative form as 
\begin{align}
  \frac{\partial \mathbf{U}}{\partial t}
    +
    \frac{\partial \mathbf{F}(\mathbf{U})}{\partial x_k}
+
\frac{\partial \mathbf{F}^v(\mathbf{U})}{\partial x_k}
 =\mathbf{S},
\end{align}
where $\mathbf{U}$ denotes the vector of conserved variables, whereas $\mathbf{F}(\mathbf{U})$ and $\mathbf{F}^{v}(\mathbf{U})$ correspond to the convective and diffusive flux vectors, respectively, while $\mathbf{S}$ represents the associated production terms. For clarity, the conservative form of the temperature tensor equation is provided in the Appendix \ref{Appendix conservative form}.
\begin{align}
    \mathbf{U}
    =
    \begin{pmatrix}
        \rho \\
        \rho v_i \\
        \frac{3+\delta}{2}\rho\theta+\frac{1}{2}\rho v^2 \\
        \frac{\delta}{2}\rho\theta_{in} \\
        \rho\Theta_{ij}+\rho v_iv_j
    \end{pmatrix},
    \qquad
    \mathbf{F(U)}
    =
    \begin{pmatrix}
        \rho v_k \\
        \rho v_i v_k+\rho\Theta_{ik} \\
        \displaystyle
        \frac{3+\delta}{2}\rho v_k\theta
        +
        \rho\Theta_{kr}v_r
        +
        \frac{1}{2}\rho v^2 v_k \\
        \displaystyle
        \frac{\delta}{2}\rho v_k\theta_{in} \\
        \displaystyle
        \rho v_k\Theta_{ij}
        +
        \rho\Theta_{ik}v_j
        +
        \rho\Theta_{jk}v_i
        +
        \rho v_i v_j v_k
    \end{pmatrix},
\end{align}
\begin{align}
      \mathbf{F}^{v}(\mathbf{U})
    =
    \begin{pmatrix}
        0 \\
        0 \\
        q_k \\
        q_k^{in} \\
        \displaystyle
        \frac{2}{5}
        \frac{
        s_i\Theta_{jk}
        +
        s_j\Theta_{ik}
        +
        s_k\Theta_{ij}}
        {\theta_{tr}}
    \end{pmatrix},
    \qquad
    \mathbf{S} =
    \begin{pmatrix}
        0 \\ 0 \\ 0 \\ \mathcal{P}^{0,1} \\ \mathcal{P}^{0,0}_{ij}
    \end{pmatrix}.
\end{align}

Here, we restrict ourselves to a one-dimensional process in the $x-$direction. The corresponding Jacobian matrix is given by
\begin{align}
\mathbf{J_c}  =
    \frac{\partial \mathbf{F}}{\partial \mathbf{U}}
=
\begin{pmatrix}
0 & 1 & 0 & 0 & 0 \\
0 & 0 & 0 & 0 & 1 \\
\frac{1}{2}v(v^2-(3+\delta)\theta-2\Theta)
&
-\frac{3v^2}{2}+\frac{1}{2}(3+\delta)\theta+\Theta
&
v
&
0
&
v
\\
-\frac{1}{2}v\delta\theta_{in}
&
\frac{\delta \theta_{in}}{2}
&
0
&
v
&
0
\\
v^3-3v\Theta
&
-3v^2+3\Theta
&
0
&
0
&
3v
\end{pmatrix}.
\end{align}
The eigenvalues of the Jacobian matrix are
\begin{align}
    \lambda
    =
    \left\{
    v\pm \sqrt{3\Theta},
    \;
    v,
    \;
    v,
    \;
    v
    \right\}.
\end{align}
Here the Jacobian matrix of the convective part has real eigenvalues,
indicating hyperbolic behavior, whereas the diffusive term is parabolic. Consequently, the 11-moment system is a hyperbolic–parabolic system.

The preceding calculation establishes hyperbolicity of the inviscid subsystem in
the realizable state space. The full system, however, is not purely hyperbolic,
because the constitutive relations introduce dissipative fluxes. Since these
dissipative terms do not act directly on all variables, the corresponding diffusion
operator is rank deficient. The model is therefore more naturally classified as a
hyperbolic--incompletely parabolic system.

This classification is consistent with the general theory of incompletely parabolic
systems arising in fluid dynamics; the compressible Navier--Stokes equations are a
standard example of such systems \cite{gustafsson1978incompletely}. More generally,
local well-posedness for quasilinear symmetric hyperbolic and hyperbolic--parabolic
systems is obtained under suitable smoothness, symmetrizability, and dissipativity
assumptions \cite{kato1975cauchy}. In partially dissipative
systems, additional coupling conditions such as the Shizuta--Kawashima condition
are often used to control undamped hyperbolic modes \cite{shizuta1985systems}.

In the present work we do not attempt to prove a complete nonlinear well-posedness
theorem for the 11-moment system. Instead, we identify the structural ingredients
needed for such an analysis: hyperbolicity of the inviscid part for realizable
states, rank-deficient parabolic dissipation, and non-negative entropy production.
A rigorous well-posedness theory for the full nonlinear system is left for future
work.

\subsection{Comparative Analysis of Models}
Compared with regularized moment models, the proposed formulation should be
viewed as a thermodynamic alternative rather than as a general replacement.
Regularized Grad-type systems, such as the R13 equations, provide an important
framework for systematically improving moment closures through order-of-magnitude
 regularization method \cite{struchtrup2003regularization}. These regularization terms introduce additional gradient-dependent dissipative terms into the Grad equations, which
substantially improve rarefaction accuracy and help remove several deficiencies of
the original Grad system, including the appearance of nonphysical sub-shocks in
shock-structure calculations \cite{struchtrup2003regularization}. The resulting
regularized systems also provide a systematic hierarchy of higher-order moment
equations with increasing rarefaction accuracy.

At the same time, hyperbolicity and realizability remain central issues for nonlinear moment systems, especially in non linear conditions \cite{cai2014globally}. Considerable an effort has therefore been devoted to globally hyperbolic regularizations and entropy-based formulations of higher-order moment equations
\cite{cai2014globally,torrilhon2010hyperbolic}.
In the linear regime, additional thermodynamic structures such as symmetric
Onsager formulations, entropy inequalities, and second-law-consistent boundary
conditions can also be constructed successfully for regularized moment systems.

The present work follows a different philosophy. Rather than constructing a large systematic hierarchy of increasingly higher-order moments, we consider an 11-moment polyatomic closure of moderate size, positioned between minimal and highly extended moment systems. This choice allows translational--internal energy exchange, entropy production, and the reversible--irreversible GENERIC structure to remain explicitly tractable. Hence, the main contribution lies not in the formal order of approximation, but in achieving a balance between tractability and physical detail, together with polyatomic energy resolution and explicit thermodynamic structure.

The choice of the number of moments of the present formulation is deliberate: increasing the number of
moments while simultaneously preserving explicit thermodynamic structure,
realizability, and tractable GENERIC coupling becomes increasingly difficult for
large nonlinear polyatomic systems.

We further compare the proposed 11-moment model with existing approaches in terms of mathematical structure, thermodynamic consistency, non-equilibrium accuracy, and limitations, as summarized in Table~\ref{tab:models_comparison}.
\begin{table}[h]
\centering
\caption{Comparison of mathematical structure, thermodynamic consistency, non-equilibrium accuracy,
and limitations of different gas dynamic models}
\scriptsize
\renewcommand{\arraystretch}{1.2}
\begin{tabular}{
|>{\centering\arraybackslash}m{1.6cm}
|>{\centering\arraybackslash}m{2.45cm}
|>{\centering\arraybackslash}m{2.2cm}
|>{\centering\arraybackslash}m{2.25cm}
|>{\centering\arraybackslash}m{2.25cm}|}
\hline
\textbf{Framework} &
\textbf{Thermodynamics} &
\textbf{Mathematical Properties} &
\textbf{Order of Accuracy} &
\textbf{Limitations} \\
\hline
NSF \cite{rahimi2016macroscopic}
&
Well-established entropy structure
&
Mixed hyperbolic--parabolic
&
First-order accuracy, $\mathcal{O}(\mathrm{Kn})$ 
&
Poor accuracy in rarefied and strongly non-equilibrium flows
\\
\hline
Grad-Type Moments \cite{struchtrup2022thermodynamically}
&
No global entropy structure in general
&
 May lose hyperbolicity and \qquad realizability
&
Second-order accuracy (near equilibrium), $\mathcal{O}(\mathrm{Kn}^2)$ 
&
Possible sub-shocks and instabilities
\\
\hline
Maximum-Entropy Closures
&
Entropy-based and symmetrizable
&
Hyperbolic in realizability  \qquad domain
&
Arbitrary-order
&
Implicit nonlinear closure relations
\\
\hline
Regularized Grad Systems (R13)  \cite{struchtrup2003regularization}
&
Second-law-consistent formulations possible
&
Improved stability and hyperbolicity
&
Third-order accuracy, $\mathcal{O}(\mathrm{Kn}^3)$ 
&
Higher mathematical complexity
\\
\hline
2T Models  \cite{aoki2020two,kumar2023h}
&
Thermodynamically admissible
&
Mixed hyperbolic--parabolic
&
First-order accuracy, $\mathcal{O}(\mathrm{Kn}^{1+\alpha})$
&
Usually limited to low-order continuum models
\\
\hline
CCR Models \cite{Anil2025CCR}
&
Explicit entropy-production structure
&
Mixed hyperbolic--parabolic
&
Second-order accuracy, $\mathcal{O}(\mathrm{Kn}^2)$ 
&
Valid mainly for moderate rarefaction
\\
\hline
Present 11-Moment Model
&
Satisfies the second law with non-negative entropy production
&
 Mixed hyperbolic--parabolic
&
 Second-order accuracy, $\mathcal{O}(\mathrm{Kn}^2)$ 
&
Extension to large nonlinear systems remains difficult
\\
\hline
\end{tabular}
\label{tab:models_comparison}
\end{table}

\section{GENERIC structure for 11-moment equations}\label{section GENERIC}
For the GENERIC structure, we choose the state variables $\mathcal{U}_\alpha$ as the mass density $\rho$, momentum density $\rho v_i$, temperature tensor $\Theta_{ij}$, and internal temperature $\theta_{in}$ as the independent fields for formulating the 11-moment equations (\ref{Conservation and balance equations}, \ref{Constitutive eqn}).

In non-equilibrium thermodynamics, energy and entropy serve as two fundamental guiding principles. For our system, the total kinetic energy of all non-interacting particles within the volume $V$ occupied by the rarefied gas can be expressed in terms of the independent system variables

\begin{align}
   E=\int_V \mathcal{E} d^3\textbf{r}=\int \rho\left(\frac{3}{2}\theta_{tr}+\frac{\delta}{2}\theta_{in}+\frac{1}{2}v^2\right) d^3\textbf{r}.
\end{align}
Here, \textbf{r} denotes the spatial position vector. The energy density gradient as
\begin{align}
    \frac{\partial \mathcal{E}}{\partial \mathcal{U}_\alpha}=\left(\frac{3}{2}\theta_{tr}+\frac{\delta}{2}\theta_{in}-\frac{1}{2}v^2,~~v_k,~~\frac{1}{2}\rho\delta_{jk},~~\rho\frac{\delta}{2}\right).
\end{align}
The entropy density is postulated to be in the form
\begin{align}
 \eta=\rho\left(\frac{1}{2}\ln [\det \Theta]-\ln\rho+\frac{\delta}{2}\ln\theta_{in}\right),
\end{align}
with entropy gradient with respect to the variables $\mathcal{U}_\alpha$ is
\begin{align}
    \frac{\partial \eta}{\partial \mathcal{U}_\alpha}=\left(\frac{\eta}{\rho}-1,~0,~\frac{1}{2}\rho \Theta_{pq}^{-1},\frac{\delta}{2}\frac{\rho}{\theta_{in}}\right).
\end{align}
\subsection{Poisson operator for reversible evolution}
A Poisson matrix $\mathcal{L}$ that adheres to the degeneracy condition and ensures the Jacobi identity holds is given by
\begin{align}
\mathcal{L} =
\scalebox{0.8}{$
    -\begin{pmatrix}
         0 & \frac{\partial}{\partial x_k}\rho & 0 & 0 \\                                                                    
         \rho \frac{\partial}{\partial x_i} & \frac{\partial}{\partial x_k}\rho v_i + \rho v_k\frac{\partial}{\partial x_i} & -\frac{\partial \Theta_{jk}}{\partial x_i} + 2\frac{\partial}{\partial x_{(j}}\Theta_{k)i} & -\frac{\partial \theta_{in}}{\partial x_i} \\
         0 & \frac{\partial \Theta_{ij}}{\partial x_k} + 2\Theta_{k(i}\frac{\partial}{\partial x_{j)}} & 0 & 0 \\
         0 & \frac{\partial \theta_{in}}{\partial x_k} & 0 & 0
    \end{pmatrix}.
$}
\end{align}
The indices enclosed in round brackets denote symmetrization. The contribution to the transport equations corresponding to the set of variables $\mathcal{U}_\alpha=(\rho, \rho v_i, \Theta_{ij},\theta_{in})_\alpha $ is given by
\begin{align}
    \mathcal{L}_{\alpha \beta}\frac{\partial \mathcal{E}}{\partial \mathcal{U}_\beta}=-
    \begin{pmatrix}
        \frac{\partial \rho v_k}{\partial x_k}\\
        \frac{\partial \rho v_i v_k}{\partial x_k}+\frac{\partial \rho \Theta_{ki}}{\partial x_k}\\
        \frac{\partial \Theta_{ij}}{\partial x_k}v_k+\Theta_{ki}\frac{\partial v_k}{\partial x_j}+\Theta_{kj}\frac{\partial v_k}{\partial x_i}\\
        v_k\frac{\partial \theta_{in}}{\partial x_k}
    \end{pmatrix}.
\end{align}
A key feature of the energy and entropy balances is that their contributions are written in divergence form
\begin{align}
    \frac{\partial \mathcal{E}}{\partial \mathcal{U}_\alpha}\mathcal{L}_{\alpha \beta}\frac{\partial \mathcal{E}}{\partial \mathcal{U}_\beta}=-\frac{\partial}{\partial x_k}\left(\frac{3+\delta}{2}\theta\rho v_k+\frac{1}{2}v^2\rho v_k+\rho\Theta_{ik}v_i\right),
\end{align}
\begin{align}
    \frac{\partial \eta}{\partial \mathcal{U}_\alpha}\mathcal{L}_{\alpha \beta}\frac{\partial \mathcal{E}}{\partial \mathcal{U}_\beta}=-\frac{\partial \eta v_k}{\partial x_k}.
\end{align}
The antisymmetric structure of the Poisson matrix \(\mathcal{L}_{\alpha\beta}\) ensures that the reversible part of the dynamics preserves both energy and entropy. Specifically, the symmetrized gradient terms involving the momentum \(\rho v_i\) and higher-order moments, such as \(\Theta_{ij}\) and \(\theta_{in}\), align with the fundamental physical symmetries and conservation laws. 

Satisfying the Jacobi identity can be complex for systems with tensorial state variables, but the specific algebraic structure of the symmetrized derivatives ensures that this identity is respected. Furthermore, the degeneracy condition \(\mathcal{L}_{\alpha\beta} \frac{\partial \eta}{\partial \mathcal{U}_\beta} = 0\) is naturally met, as the entropy functional $\eta(\mathcal{U})$ depends on scalar state variables in a way that does not affect the reversible evolution.

Importantly, the energy and entropy balances derived from this Poisson bracket are expressed in divergence form. This not only ensures local conservation but also facilitates numerical implementation and interpretation, particularly in the context of flux-based discretization schemes.
\subsection{Friction operator for irreversible evolution}
The friction matrix $\mathcal{M}$ characterizes the irreversible part of the dynamics within the GENERIC framework and is defined by
\begin{align}
    \mathcal{M}=
    \scalebox{0.8}{$
    \begin{pmatrix}
        0&0&0&0\\
        0&0&0&0\\
        0&0&M_{33}&-\frac{2}{5\rho}\frac{\partial}{\partial x_k}\left[\frac{\gamma_1 \Bar{\omega}}{\theta_{tr}}(\Theta_{jk}\Theta_{il}+\Theta_{ik}\Theta_{jl}+\Theta_{ij}\Theta_{kl})\right]\frac{\partial}{\partial x_j}\frac{2}{\delta\rho}\\
        0&0&M_{43}&-\frac{2}{\delta\rho}\frac{\partial}{\partial x_k}[\gamma_1\Bar{\omega}^2+\gamma_2]\Theta_{ik}\frac{\partial}{\partial x_i}\frac{2}{\delta\rho}
    \end{pmatrix}
    $}
\end{align}
where $M_{33}$ and $M_{43}$ defined as,
\begin{align*}
    M_{43}=-\frac{2}{5\rho}\frac{\partial}{\partial x_k}\left[\frac{\gamma_1 \Bar{\omega}}{\theta_{tr}}(\Theta_{lk}\Theta_{pq}+\Theta_{lp}\Theta_{kq}+\Theta_{lq}\Theta_{kp})\right]\frac{\partial}{\partial x_l}\frac{2}{\delta\rho},
\end{align*}
\begin{align*}
  M_{33}=\frac{2}{3}\frac{1}{\rho}[\Theta_{ir}\omega_{jr}+\Theta_{jr}\omega_{ir}]\Theta_{pq}-\frac{2}{5}\frac{\partial}{\partial x_k}\frac{\gamma_1}{\theta_{tr}^2}\frac{1}{5}\Big([\Theta_{li}\Theta_{pq}+\Theta_{lp}\Theta_{iq}+\Theta_{lq}\Theta_{ip}]\Theta_{jk}\nonumber\\+[\Theta_{lj}\Theta_{pq}+\Theta_{lp}\Theta_{jq}+\Theta_{lq}\Theta_{jp}]\Theta_{ik}+[\Theta_{lk}\Theta_{pq}+\Theta_{lp}\Theta_{kq}+\Theta_{lq}\Theta_{kp}]\Theta_{ij} \Big)\frac{\partial}{\partial x_l}\frac{2}{\rho}.
\end{align*}
\begin{align}
    \mathcal{M}_{\alpha \beta}\frac{\partial \mathcal{\eta}}{\partial \mathcal{U}_\beta}=
    \begin{pmatrix}
        0\\0\\
        -\dfrac{2\gamma_{1}}{25\theta_{\mathrm{tr}}^{2}}
\dfrac{\partial}{\partial x_k}
\Bigg\{
\Big[
\Theta_{li}\Theta_{pq}
+\Theta_{lp}\Theta_{iq}
+\Theta_{lq}\Theta_{ip}
\Big]\Theta_{jk}\\
+
\Big[
\Theta_{lj}\Theta_{pq}
+\Theta_{lp}\Theta_{jq}
+\Theta_{lq}\Theta_{jp}
\Big]\Theta_{ik}
\\
+
\Big[
\Theta_{lk}\Theta_{pq}
+\Theta_{lp}\Theta_{kq}
+\Theta_{lq}\Theta_{kp}
\Big]\Theta_{ij}
\Bigg\}
\frac{\partial\Theta^{-1}_{pq}}{\partial x_l}
\\
-\frac{2\gamma_{1}\omega}{5\theta_{\mathrm{tr}}}
\frac{\partial}{\partial x_k}
\Bigg\{
\Big(
\Theta_{il}\Theta_{jk}
+\Theta_{jl}\Theta_{ik}
+\Theta_{kl}\Theta_{ij}
\Big)
\frac{\partial\theta_{\mathrm{in}}^{-1}}{\partial x_l}
\Bigg\}\\
-\frac{2}{\delta\rho}
\frac{\partial}{\partial x_k}
\Bigg\{
\frac{\gamma_{1}\omega}{5\theta_{\mathrm{tr}}}
\Big(
\Theta_{lk}\Theta_{pq}
+\Theta_{lp}\Theta_{kq}
+\Theta_{lq}\Theta_{kp}
\Big)
\frac{\partial\Theta^{-1}_{pq}}{\partial x_l}
\\
+
\big(
\gamma_{1}\omega^{2}+\gamma_{2}
\big)
\Theta_{ik}
\frac{\partial\theta_{\mathrm{in}}^{-1}}{\partial x_i}
\Bigg\}
    \end{pmatrix}
\end{align}
Here, we use the abbreviations, $\omega_{ij}=\frac{\partial v_j}{\partial x_i}-\frac{\partial v_i}{\partial x_j}$ for simplicity\text{.}
The structure of the $\mathcal{M}_{\alpha \beta}$ matrix ensures that the irreversible contribution to the evolution equations is symmetric and positive semi-definite, as required by the second law of thermodynamics. Notably, the components of  friction matrix reflect the nonlinear couplings between the state variables $\mathcal{U}_\alpha$, consistent with non-equilibrium thermodynamic behavior. The degeneracy condition $\mathcal{M}_{\alpha \beta} \cdot \frac{\partial \mathcal{E}}{\partial \mathcal{U}_{\beta}} = 0$ is satisfied due to the structure of $\mathcal{M}$, which only couples thermodynamic fluxes such as heat flux and stress tensor rates to the entropy gradient and not to the energy gradient, thereby ensuring that irreversible processes do not alter the total energy of the system.

Now, in the limit $\delta \to 0$, the internal degrees of freedom vanish, and the conservation equations of the 11-moment model reduce to those for a monatomic gas \cite{struchtrup2005}. By considering equations \eqref{mass eqn}, \eqref{momentum eqn}, and \eqref{Conservative form of trans temp eqn}, and neglecting higher-order terms, the system reduces to the 10-moment equations reported in the literature \cite{ottinger2020formulation}. The resulting system is known to satisfy the GENERIC structure, as established in \cite{ottinger2020formulation}.

After deriving the proposed model, we test its performance by applying it to the classical one-dimensional shock structure problem in Section \ref{section shock structure}. This problem is a well-known benchmark for non-equilibrium models because it involves strong changes in flow variables and requires a correct description of both translational and internal energy. By analyzing the shock structure, we are able to evaluate the effectiveness of the model in capturing the essential non-equilibrium features of polyatomic gases.
\section{Shock structure in nitrogen gas} \label{section shock structure}
As a benchmark non-equilibrium problem, the 1-D steady shock wave structure has been analyzed using the 11-moment model.  This section aims to investigate the shock structure within nitrogen gas further. We will use the 11-moment system to compute the shock profile in nitrogen gas and evaluate the influence of nonlinear terms. For this analysis, we consider the system of governing equations given by \eqref{Conservation equation}, \eqref{internal temperature} , and \eqref{temperature tensor equation 2}, assuming that the polyatomic gas undergoes macroscopic motion primarily in the $x$-direction. The field variables, including the velocity, temperature tensor, and the translational and internal heat flux vectors, are denoted by \(v_x\), \(\Theta_{xx}\), \(q^{tr}_x\), and \(q^{in}_x\), respectively.
Furthermore, we assume that the wave is stationary with respect to the shock front. These assumptions lead to the reduction of the system described by equation in the section \ref{steady state 1D equations} to a set of ordinary differential equations, known as the shock structure equations.
\subsection{1-D shock structure computations}\label{steady state 1D equations} 
The structure of a 1-D steady shock wave links two thermodynamic equilibrium states, each defined by specific values of density ($\rho_0, \rho_1$), velocity ($v_0, v_1$), and temperature ($\theta_0, \theta_1$). The upstream state, located far ahead of the shock ($x \to -\infty$), is denoted by subscript 0, while the downstream state, located far behind the shock ($x \to \infty$), is denoted by subscript 1.

For analytical convenience and to facilitate numerical treatment, it is useful to reformulate the governing equations in a dimensionless form. This is achieved by scaling all variables with respect to their upstream reference values, leading to the introduction of the following nondimensional quantities:
\begin{equation}
\hat{\rho} = \frac{\rho}{\rho_0}, \quad
\hat{v} = \frac{v}{\sqrt{\theta_0}}, \quad
\hat{\theta} = \frac{\theta}{\theta_0}, \quad
\hat{\theta}_{xx} = \frac{\theta_{xx}}{\rho_0 \theta_0}, \quad
\hat{q} = \frac{q}{\rho_0 \sqrt{\theta_0}^3}\quad \text{and}\quad \hat{x}=\frac{x}{L}.
\end{equation}
Where $L$ is the characteristic length scale.

To investigate the shock structure in a one-dimensional framework, we consider the steady-state form of the 11-moment equations, which consist of the conservation and balance laws (\ref{Conservation and balance equations}) supplemented by the corresponding constitutive relations (\ref{Constitutive eqn}).
\begin{subequations} 
\begin{align}
    \frac{d}{d x} (\rho v_x) &= 0, \\
    \frac{d }{d x}(\rho v_x^2+\rho \theta_{tr}+\rho\theta_{xx}) &= 0, \\
    \frac{d }{d x}(c_v\rho \theta v_x+\rho \theta_{xx}v_x+\rho \theta_{tr}v_x+\frac{1}{2}\rho v^3_x+q_x) &= 0, \\
    \frac{d }{d x}\left(\frac{\delta}{2}\rho \theta_{in}v_x+q_x^{in}\right) &= \mathcal{P}^{0,1}, \\
    \frac{d}{dx}\rho \theta_{xx}v+\frac{4}{3}\rho(\theta_{tr}+\theta_{xx})\frac{d v}{d x}+\frac{2}{3}\frac{d}{d x}\frac{4\theta_{tr}+7\theta_{xx}}{5\theta_{tr}+2\theta_{xx}}q^{tr} &= \mathcal{P}_{\langle i j\rangle}^{0,0}.
\end{align}
\label{Conservation equation in 1D}
\end{subequations}
Constitutive equation,\\
\begin{align}
    q^{tr} &= -\frac{5\theta_{tr}+2\theta_{xx}}{5\theta_{tr}} \Bigg( 
\frac{\zeta_{11}}{5\theta_{tr}^2} \frac{1}{2\theta_{tr}-\theta_{xx}} 
\Big[ (10\theta_{tr}+\theta_{xx})\frac{d \theta_{tr}}{\partial x} \nonumber \\ 
&\quad + (4\theta_{tr}-5\theta_{xx})\frac{d \theta_{xx}}{d x} \Big] 
+ \frac{\zeta_{12}}{\theta_{tr}\theta_{in}^2} 
\Big[(\theta_{tr}+\theta_{xx})\frac{d \theta_{in}}{d x}\Big] 
\Bigg),
\end{align}
\begin{align}
     q^{in}=-\frac{\zeta_{12}}{5\theta_{tr}^{2}} \left[  \frac{10\theta_{tr}+\theta_{xx}}{2\theta_{tr}-\theta_{xx}}\frac{d \theta_{tr}}{d x}+\frac{4\theta_{tr}-5\theta_{xx}}{2\theta_{tr}-\theta_{xx}}\frac{d \theta_{xx}}{d x}\right]-  \frac{\zeta_{22}}{\theta_{tr}\theta_{in}^2}\left[(\theta_{tr}+\theta_{xx})\frac{d \theta_{in}}{d x}\right].
\end{align}
In this work, we adopt a Prandtl number $ \text{Pr} = \frac{14}{19} $ \cite{elizarova2007numerical}, which closely matches values obtained from DSMC simulations and experimental data for nitrogen gas. The shear viscosity $ \mu $ is depicted as   temperature-dependent, using the relation  
 $\mu = \mu_0 \left( \frac{T}{T_0} \right)^\nu,$   
where $ \mu_0 $ and $ T_0 $ are reference values and $ \nu=0.78 $ is the viscosity exponent \cite{jiang2019computation}. The bulk viscosity $ \mu_b $ is taken as 73\% of the shear viscosity \cite{kosuge2016shock}, in agreement with DSMC-based models and the approach described by Bird. It is worth noting that some authors instead adopt a value of 80\% for the bulk viscosity relative to the shear viscosity \cite{song2025non}.
The Prandtl number is determined by the ratio of shear viscosity over heat conductivity $\kappa$ \cite{elizarova2007numerical,rahimi2016macroscopic}
\begin{align}
    \text{Pr} = \frac{5+\delta}{2}\frac{R}{\kappa}\mu .
\end{align}
Here, $\kappa=\kappa(\zeta_{11},\zeta_{12},\zeta_{22})$ \eqref{Reduced NSF constitutive eqn}. Hence, the Prandtl number is determined by the choice of the dimensionless moment of inertia parameter $\varsigma$. For the numerical simulations presented in this work, we use $\varsigma=0.442$, which yields the corresponding Prandtl number stated above.


\subsection{Numerical methodology}
To solve the system of equations (\ref{Conservation equation in 1D}), we adopt the finite difference and midpoint method as a numerical strategy. As a first step, the number of variables in the system is reduced to simplify the computational procedure.
 This is achieved by integrating the first three equations, starting from a point before the shock, to derive expressions for density, traceless part of stress $\theta_{xx}$, and translational heat flux in terms of rest variables. After that, we employ the midpoint method \cite{ascher1998computer} to solve the system of ordinary differential equations for the remaining variables \cite{kumar2024capturing}.

In the dimensionless formulation, the upstream (pre-shock) conditions are set as $\rho_0 = 1$, $v_0 = \sqrt{\gamma}\, M_0$, and $\theta_0 = 1$, with the dynamic temperature $\vartheta = 0$. At equilibrium, the conditions $\theta_{xx} = 0$, $q^{tr} = 0$, $q^{in} = 0$, and $\vartheta = 0$ are imposed. To analyze the shock structure, we integrate the first three steady-state equations from the system (\ref{Conservation equation in 1D}) in the upstream region. This yields a set of algebraic relations that express the density, stress,  and translational heat flux in terms of the velocity $v$, temperature $\theta$, internal heat flux $q^{in}$ and dynamic temperature $\vartheta$:
\begin{subequations}
    \begin{align}
\rho=\sqrt{\gamma}\frac{M_0}{v},\\
    \theta_{xx}=\frac{v}{\sqrt{\gamma}M_0}+\sqrt{\gamma}M_0v-v^2-\theta-\vartheta,\\
    q^{tr}=\frac{1}{2}\Big(-2q^{in}+M_0 v^2 \sqrt{\gamma}-2v(1+M_0^2\gamma)+M_0\sqrt{\gamma}(5+M_0^2\gamma+\delta-3\theta-\delta\theta)\Big).
\end{align}
\end{subequations}
Where $\gamma$ represents the ratio of specific heats, defined as $\gamma=(5+\delta)/(3+\delta)$ and for nitrogen gas, its value is approximately 1.4 \cite{elizarova2007numerical}. The Mach number $M_0$ denotes, representing the ratio between the upstream flow velocity (relative to the shock) and the corresponding speed of sound.
  
Upon solving the system of ordinary differential equations, the gas transitions to equilibrium states both upstream and downstream of the shock. To ensure that the problem is well-posed, suitable boundary conditions must be imposed. These conditions are determined using the Rankine–Hugoniot relations, which establish the conservation-based connections between the equilibrium states on either side of the shock. The resulting formulation requires four boundary conditions to fully specify the solution, which are as follows
\begin{align}
        v_0(x\rightarrow-L)=\sqrt{\frac{5}{3}}M_0,~~v_1(x\rightarrow L)=\frac{5+M_0^2\gamma+\delta}{M_0\sqrt{\gamma}(4+\delta)},\nonumber\\ \theta_0(x\rightarrow-L)=1,\quad
       \theta_1(x\rightarrow L)=\frac{(5+M_0^2\gamma+\delta)(3M^2_0\gamma+M_0^2\delta\gamma-1)}{M_0^2\gamma(4+\delta)^2}.\label{Boundary_condition}
\end{align}
These boundary conditions serve as critical inputs for numerically solving the system of ordinary differential equations governing the shock structure. They ensure a smooth transition from the upstream to downstream equilibrium states and help enforce the conservation of mass, momentum, and energy across the shock.

Now, we apply the midpoint finite difference method \cite{ascher1998computer} to numerically solve the coupled system of ordinary differential equations governing velocity, translational and internal temperature, temperature tensor, and translational and internal heat flux. The method provides second-order accuracy and is particularly effective for capturing smooth but rapid transitions, such as those present in shock structures.

As an initial guess, the profiles of velocity and temperatures are initialized using hyperbolic tangent ($\tanh(x)$) functions, which approximate the steep but continuous variations expected across the shock layer. The computational domain is taken as $[-L, L]$, where $L$ is sufficiently large to ensure that upstream and downstream equilibrium states are well-resolved. This domain is uniformly discretized into $N + 1$ grid points, denoted $x_i$ for $i = 0, 1, ..., N$, with uniform step size $h$.

The midpoint scheme approximates spatial derivatives and field values at cell interfaces as follows:
\begin{align}
    \left. \frac{dy}{dx} \right|_{i-1/2} \approx \frac{y_i - y_{i-1}}{h}, \quad y|_{i-1/2} \approx \frac{y_i + y_{i-1}}{2}, \quad i = 1, 2, \dots, N.
\end{align}
To evaluate the system at interior nodes $x_1$ through $x_{N-1}$, field values at the ghost points $x_0$ and $x_N$ are required. These are provided through the boundary conditions stated in equation (\ref{Boundary_condition}). As a result, the discretization leads to a system of $4(N + 1)$ nonlinear algebraic equations for the $4(N + 1)$ unknowns: velocity $v$, total temperature $\theta$, and dynamic temperature $\vartheta$, and internal heat flux $q^{in}$, all defined on the full mesh (including boundaries).\\
In this study, all numerical computations were carried out in MATLAB\textsuperscript{©}. To solve the resulting nonlinear system of equations, we employed the Powell dogleg method, following the approach outlined in \cite{powell1968fortran}. Although the second-order numerical scheme used here may face convergence issues if the computational domain is overly large, we found that choosing an appropriate domain size leads to stable and highly accurate results. In our simulations, for different Mach numbers, the numerical solver consistently reached an approximate relative accuracy of $10^{-24}$ in around 12 iterations. The shock structures of the proposed model were computed using $N = 2500$ uniformly spaced grid points.
\subsection{Results and discussion}
In this section, we solve the present model for nitrogen gas and compare the resulting profiles with DSMC simulations \cite{cai2014nrxx} and experimental observations \cite{alsmeyer1976density}, considering Mach numbers up to 3.8. In the numerical solution of the equations, the origin of the $x/\lambda_0$ axis is selected such that the density at $x/\lambda_0=0$ corresponds to the average of the upstream and downstream density values.

To facilitate comparison under varying flow conditions, it is useful to introduce dimensionless variables. Specifically, the density and temperature are normalized by their respective upstream values as defined below:
\begin{equation}
    \hat{\rho} = \frac{\rho - \rho_0}{\rho_1 - \rho_0},~~ \hat{\theta} = \frac{\theta - \theta_0}{\theta_1 - \theta_0}, ~~\hat{\theta}_{tr} = \frac{\theta_{tr} - \theta_0}{\theta_1 - \theta_0},  ~~\text{and} ~~ \hat{\theta}_{in} = \frac{\theta_{in} - \theta_0}{\theta_1 - \theta_0},
\end{equation}
where, the subscript 0 denotes upstream conditions (before the shock), while the subscript 1 refers to downstream conditions (after the shock).

\begin{figure}[htbp]
    \centering
    \includegraphics[width=1.0\textwidth]{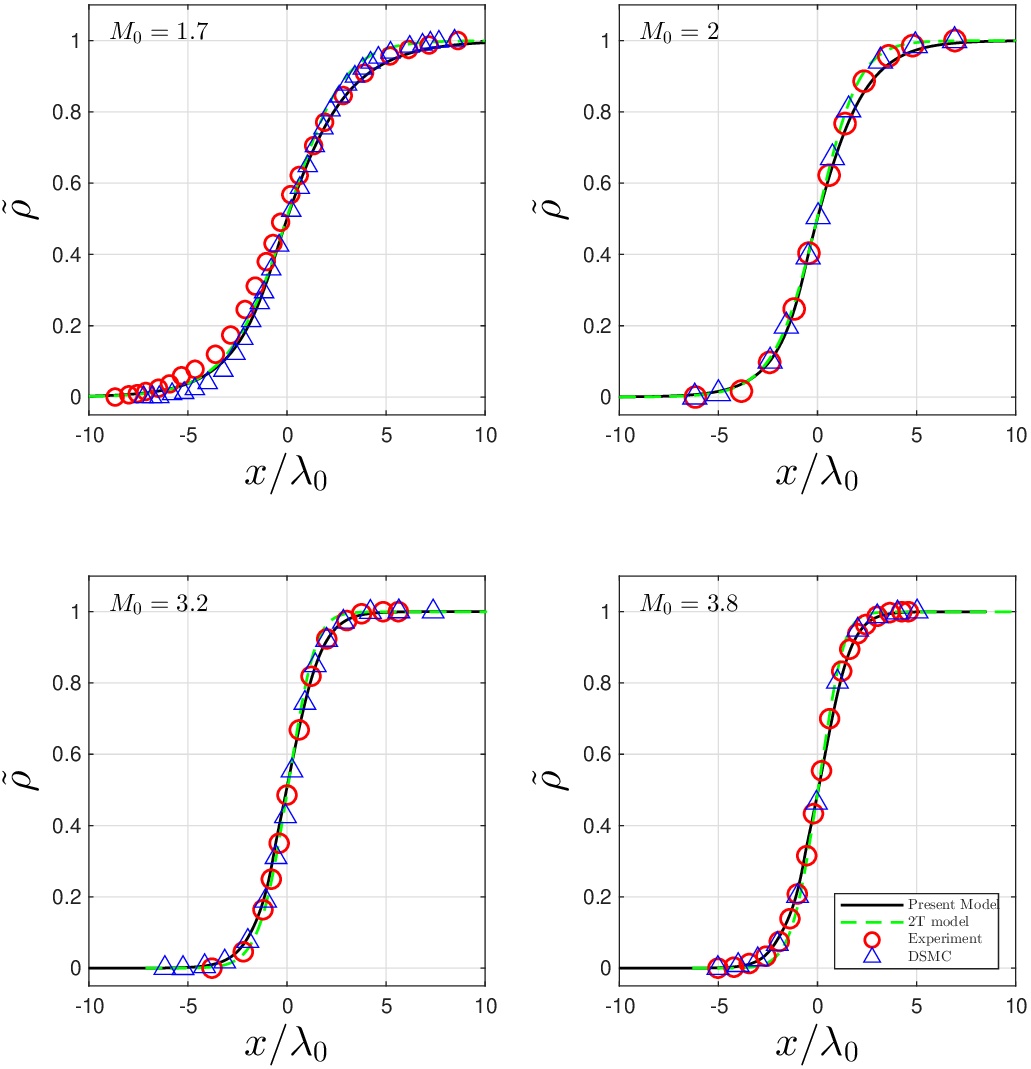}
    \caption{Normalized density profiles for various Mach numbers. Solid black curves denote the present model, dashed green curves denote the $2$T model, while symbols represent the experimental data (\textcolor{red}{$\circ$}), DSMC data (\textcolor{blue}{$\triangle$}).}
    \label{fig:normalized_density}
\end{figure}

\begin{figure}[htbp]
    \centering
    \includegraphics[width=1.0\textwidth]{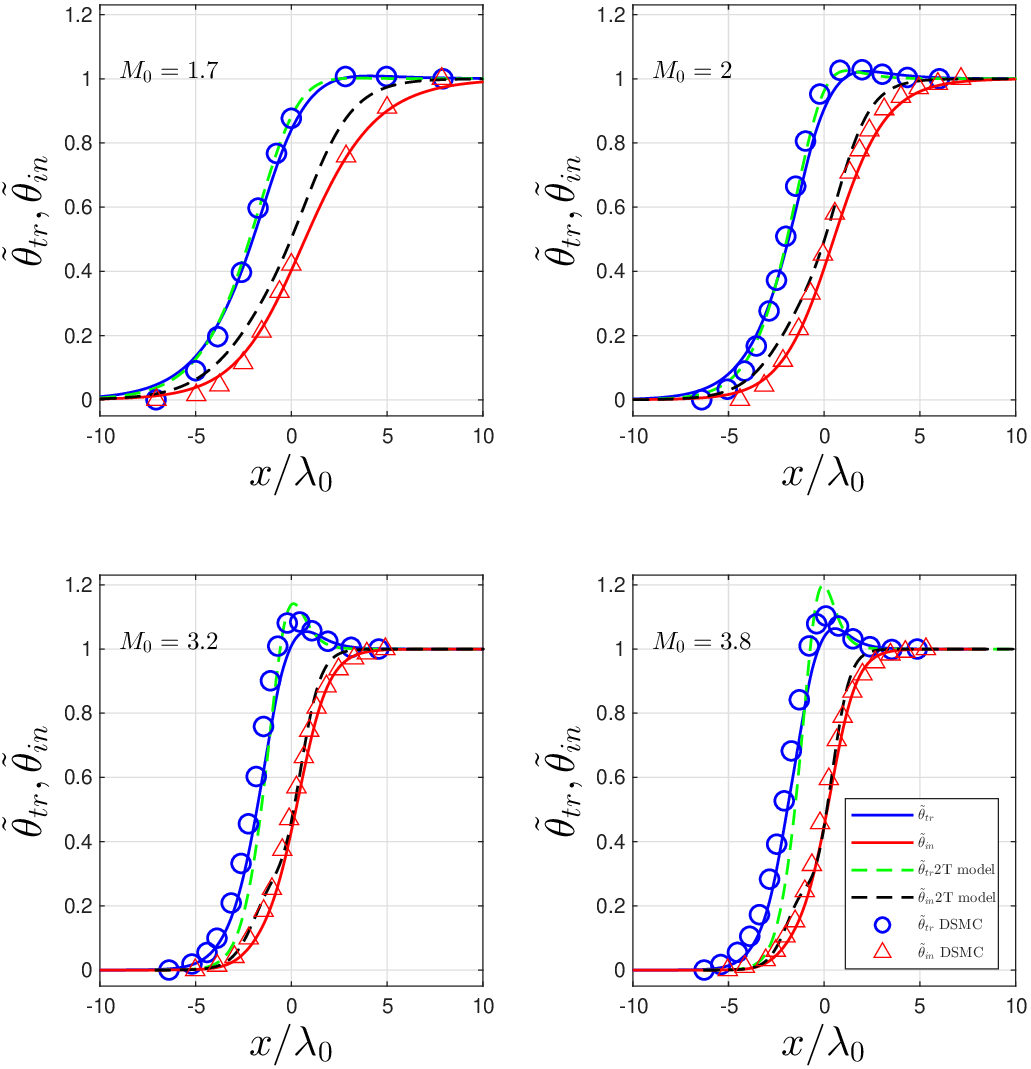}
    \caption{Profiles of the normalized translational temperature $\tilde{\theta}_{tr}$ and internal temperature $\tilde{\theta}_{in}$ for various Mach numbers. Solid curves denote the present model $(\text{blue: } \tilde{\theta}_{tr},\ \text{red: } \tilde{\theta}_{in})$, dashed curves denote the $2$T model $(\text{green: } \tilde{\theta}_{tr},\ \text{black: } \tilde{\theta}_{in})$, and symbols represent DSMC data.
}
    \label{fig:normalized_temp}
\end{figure}

Figures \ref{fig:normalized_density} and \ref{fig:normalized_temp} show the normalized profiles of density and temperature as functions of $x/\lambda_0$ for different Mach numbers. In Figure \ref{fig:normalized_density}, the numerical results for the density profile within the shock wave in nitrogen gas are compared with experimental and DSMC data for Mach numbers ranging from 1.7 to 3.8. The present system demonstrates strong agreement with both experimental and DSMC results across these Mach numbers, particularly in the normalized density. The solid black curves represent the predictions of the present 11-moment model, while the dashed green curves correspond to the 2T model \cite{kumar2024capturing}. The symbols represent the experimental data \cite{alsmeyer1976density} and the DSMC results \cite{cai2014nrxx}. The present system demonstrates strong agreement with both experimental and DSMC results across these Mach numbers, particularly in the normalized density.

Figure \ref{fig:normalized_temp} presents a comparison of the translational and internal temperature profiles with DSMC data for Mach numbers ranging from 1.7 to 3.8. The solid blue and red curves represent the translational temperature $\theta_{tr}$, and internal temperature $\theta_{in}$, respectively, obtained from the present 11-moment model. The dashed green and dashed black curves correspond to the predictions of the 2T model \cite{kumar2024capturing} for $\theta_{tr}$, and $\theta_{in}$, respectively. The symbols denote the DSMC data reported in \cite{cai2014nrxx}. While the 2T model shows noticeable deviations in the temperature profiles, particularly in the non-equilibrium region, the present 11-moment model demonstrates much better agreement with the DSMC data. A strong agreement between the results from the present model and the DSMC data is observed, although a small discrepancy in the translational temperature is noted in the downstream region at Mach numbers 3.2 and 3.8.

At higher Mach numbers, particularly beyond Mach number 4, the agreement with DSMC results begins to deteriorate slightly, especially in the downstream translational temperature profiles. This deviation may be attributed to the assumption of a constant relaxation parameter \(\delta\), which implies fixed specific heat capacities. While adequate at moderate temperatures, this assumption becomes increasingly restrictive under strong temperature gradients, as occurs in high-speed flows, where energy redistribution among molecular modes becomes temperature dependent.

\section{\label{Section Summary and future direction}Summary and future directions}
In this work, we derived a system of 11-moment equations to describe the non-equilibrium behavior of polyatomic gases. The formulation accounts for both translational and internal degrees of freedom, leading to a richer description of non-equilibrium thermodynamic processes.
We have developed a thermodynamically consistent 11-moment model for polyatomic gases that adheres fully to the GENERIC framework. The strength of the model lies in its structural rigor: reversible and irreversible processes are governed by Poisson and friction matrices, respectively, and their formulation ensures automatic satisfaction of the second law of thermodynamics. Closure relations and entropy-production terms are derived from thermodynamic first principles, making explicit the entropy density, entropy flux, entropy-production mechanisms, and translational--internal energy coupling that characterize gas dynamics away from equilibrium.

The main contribution of the present work is not to claim that the GENERIC framework provides universally improved predictions over existing models, but rather to demonstrate that the proposed 11-moment model can be formulated within a transparent and thermodynamically consistent structure. The GENERIC framework explicitly identifies the reversible and irreversible contributions, entropy flux, entropy production, and translational–internal energy coupling in nonequilibrium gas dynamics. Moreover, a system formulated within the GENERIC framework automatically satisfies the conservation laws and is consistent with the second law of thermodynamics through its built-in entropy-production structure. Any agreement with DSMC or experimental data should therefore be attributed to the specific 11-moment model developed within this framework, rather than to the GENERIC structure itself.

To test the model’s capabilities, we applied it to compute shock structures in nitrogen gas. The results showed excellent agreement with experimental and DSMC data up to Mach 3.8, confirming that the framework effectively captures non-equilibrium phenomena in rarefied conditions.

Looking ahead, the assumption of a constant relaxation parameter $\delta$ corresponding to temperature independent heat capacities may limit the accuracy of the model in high-temperature or high-Mach regimes. A promising direction for future work is to incorporate a temperature dependent $\delta$ while preserving the GENERIC structure, which could further enhance the model's predictive capabilities in extreme flow environments.

 \appendix

 \section{Derivation} \label{Appendix for derivation1}
The production terms $\mathcal{P}^{0,1}$ and $\mathcal{P}_{ij}^{0,0}$ are non-negative and $\varrho_{\{ijk\}}^{0,0}=0$. The entropy production remains non-negative, for this the equation~\eqref{Entropy equation} can be written as
\begin{align}
\rho \frac{Ds }{Dt}+\frac{\partial \left( \frac{q_{k}^{in}}{\theta _{in}}+
\frac{s_{k}}{\theta _{tr}}\right) }{\partial x_{k}} =q_{k}^{in}\frac{\partial \theta _{in}^{-1}}{\partial x_{k}}
+\frac{1}{5\theta _{tr}}s_{i}\Theta _{i\alpha }^{-1}\left[ \Theta _{l\alpha }\Theta
_{jk}+\Theta _{lj}\Theta _{\alpha k}+\Theta _{lk}\Theta _{\alpha j}\right] 
\frac{\partial \Theta _{jk}^{-1}}{\partial x_{l}}.
\end{align}
The equations are rewritten in an equivalent form without loss of generality. To capture the interaction between translational and internal energy modes, we introduce the coupling parameter $\bar{\omega}$, which couples the corresponding heat fluxes. This coupling is essential for describing the energy exchange between the two modes and hence the non-equilibrium physics of polyatomic gases more realistically. In the limit $\bar{\omega}=0$, the coupling vanishes, the heat fluxes become decoupled, and the physical richness of the model is significantly reduced.
\begin{subequations}
   \begin{align}
\rho \frac{Ds }{Dt}+\frac{\partial \left( \frac{q_{k}^{in}}{\theta _{in}}+
\frac{s_{k}}{\theta _{tr}}\right) }{\partial x_{k}} =(q_{i}^{in}-\bar{w} s_i) \Theta_{i\alpha}^{-1} \Theta_{\alpha k} \frac{\partial \theta_{in}^{-1}}{\partial x_k}\nonumber\\
+s_i \Theta_{i\alpha}^{-1}\left(\frac{1}{5\theta _{tr}}\left[ \Theta _{l\alpha }\Theta
_{jk}+\Theta _{lj}\Theta _{\alpha k}+\Theta _{lk}\Theta _{\alpha j}\right] 
\frac{\partial \Theta _{jk}^{-1}}{\partial x_{l}}+\bar{w}  \Theta_{\alpha k} \frac{\partial \theta_{in}^{-1}}{\partial x_k}\right), 
\end{align}
\end{subequations}
for entropy production to be non-negative, then we take the equation \eqref{eqn of si} with 
\begin{subequations}
    \begin{align}
    &q_i^{in} - \bar{\omega} s_i = \gamma_2 \Theta_{ik} \frac{\partial \theta_{in}^{-1}}{\partial x_k}.\label{eqn of qin with const}
\end{align}
\end{subequations}
Substituting the expression of $s_i$ from \eqref{eqn of si} into \eqref{eqn of qin with const}, we obtained internal heat flux  \eqref{Qin}. After, substituting the expression for $s_i$ from \eqref{eqn of si}  into \eqref{Qtr} and taking the trace part, we get
\begin{subequations}
    \begin{align}
    q_i^{tr} &= -\frac{\gamma_{1}}{\theta_{tr}} \frac{\partial \theta_{tr}}{\partial x_i} -\gamma_1 \bar{\omega} \frac{\theta_{tr}}{\theta_{in}^2} \frac{\partial \theta_{in}}{\partial x_i},\\
    q_i^{in}& = -\frac{\gamma_{1} \bar{\omega}}{\theta_{tr}} \frac{\partial \theta_{tr}}{\partial x_i} -(\gamma_1 \bar{\omega}^2 + \gamma_2) \frac{\theta_{tr}}{\theta_{in}^2} \frac{\partial \theta_{in}}{\partial x_i}.
\end{align} \label{trace part of eqn qtr qin}
\end{subequations}
\\
\section{Conservative form} \label{Appendix conservative form}
We have temperature tensor equation in non conservative form as \eqref{translational temp eqn}
\begin{subequations}
    \begin{align}
    \rho \frac{\partial \Theta_{ij}}{\partial t}+\rho v_k \frac{\partial \Theta_{ij}}{\partial x_k}+\frac{\partial(\rho \Theta_{ik}v_j)}{\partial x_k}-v_j\frac{\partial (\rho \Theta_{ik})}{\partial x_k}+\frac{\partial(\rho \Theta_{jk}v_i)}{\partial x_k}\nonumber\\-v_i\frac{\partial (\rho \Theta_{jk})}{\partial x_k}+\frac{2}{5}\frac{\partial \left(\frac{s_i\Theta_{jk}+s_j\Theta_{ik}+s_k\Theta_{ij}}{\theta^{tr}}\right)}{\partial x_k}= \mathcal{P}_{ij}^{0,0},\\
   \frac{\partial \rho \Theta_{ij}}{\partial t}-\Theta_{ij}\frac{\partial \rho}{\partial t} +\frac{\partial \rho v_k\Theta_{ij}}{\partial x_k}-\Theta_{ij}\frac{\partial \rho v_k}{\partial x_k}+\frac{\partial(\rho \Theta_{ik}v_j)}{\partial x_k}+\frac{\partial(\rho \Theta_{jk}v_i)}{\partial x_k}\nonumber\\+\frac{2}{5}\frac{\partial \left(\frac{s_i\Theta_{jk}+s_j\Theta_{ik}+s_k\Theta_{ij}}{\theta^{tr}}\right)}{\partial x_k}= \mathcal{P}_{ij}^{0,0}+\underbrace{v_j\frac{\partial (\rho \Theta_{ik})}{\partial x_k}+v_i\frac{\partial (\rho \Theta_{jk})}{\partial x_k}}\label{change to conservative form},
\end{align}
\end{subequations}
We now evaluate only the under braces terms in \eqref{change to conservative form}. Using \eqref{momentum eqn} in under braces term, we obtain
\begin{subequations}
    \begin{align}
   = -v_j\left[\frac{\partial \rho v_i}{\partial t}+\frac{\partial \rho v_i v_k}{\partial x_k}\right]-v_i\left[\frac{\partial \rho v_j}{\partial t}+\frac{\partial \rho v_j v_k}{\partial x_k}\right],\\
 =   -v_j\left[v_i\frac{\partial \rho }{\partial t}+\rho\frac{\partial  v_i}{\partial t}+v_i\frac{\partial \rho  v_k}{\partial x_k}+ \rho v_k\frac{\partial v_i}{\partial x_k}\right]-v_i\bigg[v_j\frac{\partial \rho }{\partial t}+\rho\frac{\partial  v_j}{\partial t}+v_j\frac{\partial \rho  v_k}{\partial x_k}\nonumber\\+ \rho v_k\frac{\partial v_j}{\partial x_k}\bigg]\label{Derivation3},
\end{align}
\end{subequations}
by using \eqref{mass eqn} in \eqref{Derivation3}, we get
\begin{subequations}
     \begin{align}
  = -v_j\left[\rho\frac{\partial  v_i}{\partial t}+ \rho v_k\frac{\partial v_i}{\partial x_k}\right]-v_i\left[\rho\frac{\partial  v_j}{\partial t}+ \rho v_k\frac{\partial v_j}{\partial x_k}\right],  \\
 =- \frac{\partial \rho v_i v_j}{\partial t} + v_i v_j\frac{\partial \rho }{\partial t}-\frac{\partial \rho v_i v_j v_k}{\partial x_k}+v_i v_j\frac{\partial \rho v_k}{\partial x_k} \label{Derivation4},
\end{align}
\end{subequations}
again by using \eqref{mass eqn} in \eqref{Derivation4}, we get
\begin{align}
= - \frac{\partial \rho v_i v_j}{\partial t} -\frac{\partial \rho v_i v_j v_k}{\partial x_k}\label{Derivation2},
\end{align}
now in place of under braces term  in
\eqref{change to conservative form}, we use \eqref{Derivation2}, then
after solving, we obtain temperature tensor equation in conservative form
\begin{align}
    \frac{\partial}{\partial t}(\rho \Theta_{ij}+\rho v_i v_j)+  \frac{\partial}{\partial x_k}(\rho v_k \Theta_{ij}+\rho \Theta_{ik}v_j+\rho \Theta_{jk}v_i+\rho v_iv_jv_k)\nonumber\\+ \frac{2}{5} \frac{\partial}{\partial x_k}\left(\frac{s_i\Theta_{jk}+s_j\Theta_{ik}+s_k\Theta_{ij}}{\theta_{tr}}\right)=\mathcal{P}_{ij}^{0,0}\label{Conservative form of trans temp eqn}
\end{align}

\section*{Acknowledgments}
MA gratefully acknowledges the financial support provided by the Birla Institute of Technology and Science, Pilani, Pilani campus. ASR acknowledges the financial support from the Science and Engineering Research Board under Grant No. MTR/2021/000417.


\bibliographystyle{abbrv}
\bibliography{references}
\end{document}